\documentclass[pre,twocolumn,superscriptaddress,preprintnumbers,amsmath,amssymb]{revtex4-1}

\usepackage{epsfig}
\usepackage{amsmath}
\usepackage{amssymb}
\usepackage{amsfonts}
\usepackage{mathptmx}
\usepackage{dcolumn}
\usepackage{eucal}
\usepackage{bm,bbm}
\usepackage{color}
\usepackage[colorlinks,linkcolor=blue,citecolor=blue]{hyperref}
\usepackage{epstopdf}
\usepackage{bbold}
\usepackage{url}

\usepackage{dsfont}

\newcommand{\Trace}{{\rm Tr}}
\renewcommand{\d}{{\rm d}}

\def \ket#1{\mathinner{|{#1}\rangle}}
\def \bra#1{\mathinner{\langle{#1}|}}

\newcommand{\braket}[2]{{\mathinner{\langle {#1} | {#2} \rangle}} }
\newcommand{\matrixel}[3]{{\mathinner{\langle{#1}| {#2} | {#3}\rangle}} }
\newcommand{\average}[1]{\langle #1 \rangle}
\newcommand{\Forder}{\overrightarrow{\mathcal T}}
\newcommand{\GFunc}{\mathcal{G}}
\newcommand{\Prob}{\mathcal{P}}

\begin{document}

\title{Measurement-dependent corrections to work distributions arising from quantum coherences} 

\author{P. Solinas}
\email[]{paolo.solinas@spin.cnr.it}
\affiliation{SPIN-CNR, Via Dodecaneso 33, 16146 Genova, Italy.}
\author{H. J. D. Miller}
\affiliation{CEMPS, Physics and Astronomy, University of Exeter, Exeter EX4 4QL, United Kingdom.}
\author{J. Anders}
\affiliation{CEMPS, Physics and Astronomy, University of Exeter, Exeter EX4 4QL, United Kingdom.}

\date{\today}

\begin{abstract}
For a quantum system undergoing a unitary process work is commonly defined based on the Two Projective Measurement (TPM) protocol which measures the energies of the system before and after the process. However, it is well known that  projective measurements disregard quantum coherences of the system with respect to the energy basis, thus removing potential quantum signatures in the work distribution. Here we consider weak measurements of the system's energy difference and establish corrections to work averages arising from initial system coherences. We discuss two weak measurement protocols that couple the system to a detector, prepared and measured either in the momentum or the position eigenstates. Work averages are derived for when the system starts in the proper thermal state versus when the initial system state is a pure state with thermal diagonal elements and coherences characterised by a set of phases. We show that by controlling only the phase differences between the energy eigenstate contributions in the system's initial pure state, the average work done during the same unitary process can be controlled. By changing the phases alone one can toggle from regimes where the systems absorbs energy, i.e. a work cost, to the ones where it emits energy, i.e. work can be drawn. This suggests that the coherences are additional resources that can be used to manipulate or store energy in a quantum system.
\end{abstract}

\date{\today}

\maketitle

\section{Introduction}

The concept of mechanical work is a cornerstone at the basis of classical physics. When a closed system is displaced by an external force, the force does work on the system and its energy changes. In this case, the work corresponds to the energy supplied to the system. Surprisingly, these ideas have not been discussed in the framework of quantum mechanics until a few years ago \cite{campisi2011colloquium,campisi2011erratum,esposito2009Erratum,esposito2009nonequilibrium}. The main reason for this lack is that work cannot be associated to a hermitian operator \cite{talkner2007fluctuation} - for a closed system, i.e., not in contact with a heat bath, work is equivalent to the change of the internal energy which is non-local in time. To by-pass this problem, the first proposals for the measurement of  work at the quantum level were based on the Two Projective Measurement (TPM) protocol \cite{campisi2011colloquium,campisi2011erratum,esposito2009Erratum,esposito2009nonequilibrium,kurchan2000quantum,tasaki2000jarzynski}. The energy of the system is projectively measured at the beginning and at the end of the evolution in order to extract information about the moments of the work done. However, due to the first projective measurement, the system collapses in an eigenstate of the initial Hamiltonian and the initial coherences between eigenstates of the initial Hamiltonian are destroyed. This changes the dynamics of the system and cancels, in Feynman's words, interfering alternatives \cite{feynman1965quantum, Cotler2016}.

The effect of the projective measurement on the work statistics has been underestimated because the problem of defining work at the quantum level has mostly been considered within the context of quantum thermodynamics and fluctuation theorems \cite{Jarzynski1997Nonequilibrium,kurchan2000quantum,tasaki2000jarzynski, talkner2007fluctuation, campisi2011colloquium, campisi2011erratum, esposito2009Erratum, esposito2009nonequilibrium, Koski2013NatPhys, Batalho2014}. Here the system is customarily assumed to start in a thermal state, with no coherences between the eigenstates of the initial Hamiltonian.  In this situation, the first energy measurement has indeed no effect and the work distribution is well described as a statistical ensemble of iterative experiments.

Only recently there has been a growing awareness of the limitation of the projective measurement approach \cite{solinas2013work, gasparinetti2014heat, roncaglia2014work, solinas2015fulldistribution,Elouard2015stochastic, talkner2016, Kammerlander2016Coherence, solinas2016probing, Perarnau-Llobet2017No-Go, lostaglio2017quantum}.
This interest is fuelled by the growing evidence that quantum effects and coherences are believed to be a fundamental quantum resource \cite{Baumgratz2014QuantCoher, Winter2016TheoCoherences, liu2016theory, Napoli2016, Deffner2016, Korzekwa2016, Perarnau-Llobet2017No-Go, Elouard2017, miller2016time, lostaglio2017quantum}; they play an important role not only in quantum information (computation and cryptography) \cite{nielsen-chuang_book} but also in quantum biology \cite{Engel2007, mohseni2008environment, Collini2010, Panitchayangkoon2010, Lloyd2011,OReilly2014, kolchinsky2017maximizing}. Within the thermodynamics context it is natural to ask if the presence of quantum coherences modifies the quantum energy exchange.

A number of proposals have been put forward to characterize and measure the energy exchange at the quantum level \cite{roncaglia2014work, solinas2015fulldistribution, talkner2016, solinas2016probing}. Here we show that the quantum work distribution generated when an external drive acts on a system, depends on, and can be controlled by, the initial coherences of the system. We discuss how these two quantities can be measured with two complementary protocols \cite{solinas2016probing}. The first employs a quantum detector coupled to the system with varying coupling strengths, which allows one to measure the work characteristic function. The second protocol implements a more traditional pointer measurement scheme which determines directly the work distribution. 
To exemplify the peculiarities of the resulting work distributions we here focus on evaluating the average exponentiated work and the average work. The Jarzynski Equality (JE) \cite{Jarzynski1997Nonequilibrium, talkner2007fluctuation} for the average exponentiated work is a powerful tool since for initial thermal states it holds for any initial temperature and any drive when the TPM is used. The JE allows insight in regimes in which the system is driven out-of-equilibrium that are challenging to characterise otherwise. Here we use these features to study the effect of coherences on the energy exchange since any deviation from JE can be immediately traced back to the presence of initial coherences.  
We will also discuss how the average work, i.e. the first moment of the work distribution, is influenced by and can be controlled through quantum coherences.

We find that the two protocols give different work distributions, different  deviations from the JE and values of the average work. Despite the quantitative differences in the two protocols, both of them clearly signal the effect of quantum coherences. In particular, the average work can be positive (the system absorbs energy from the drive) or negative (the system emits energy into the drive) confirming that the quantum coherences may be exploited as energetic resources.

\section{System and detector dynamics}

\subsection{Time-dependent system and detector Hamiltonian}

We consider a closed quantum system, $S$, driven by an external, classical field, so that its dynamics is governed by the time-dependent Hamiltonian $\hat H_S(t)$ for times $t \in [0, \tau]$. The system does not interact with any environment and cannot exchange heat. The work done on the system, $W$, by the external driving field is thus associated with the system's energy change. 

\begin{figure}[t]
\includegraphics[trim=3cm 8.8cm 3cm 4cm,  clip, width=0.48\textwidth]{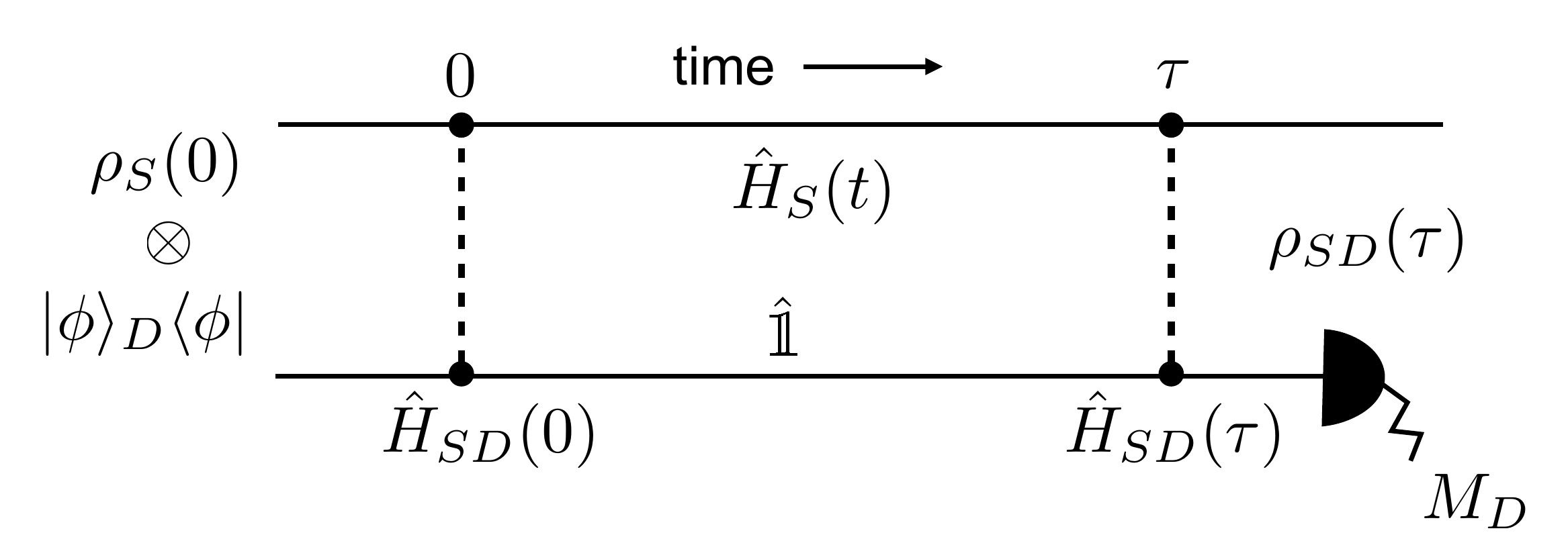}
\caption{\label{fig:S+D} System $S$ and detector $D$ are initially uncorrelated in a state $\rho_S (0) \otimes \ket{\phi}_D\bra{\phi}$. The global unitary \eqref{eq:V} couples them at time 0, and then evolves the system alone with the time-dependent Hamiltonian $\hat H_S(t)$ while not affecting the detector. $S$ and $D$ are coupled again at time $\tau$ resulting in the final joint state $\rho_{SD}(\tau)$. The measurement $M_D$ of the detector then reads out the work done on the system during its evolution between times $0$ and $\tau$. Here we discuss two variations of the initial detector state and its final measurement, Protocols 1 and 2, and find that work one associates with the system depends on this choice.}
\end{figure}

To measure the system's energy change \emph{weakly} the system will be coupled to a quantum detector, $D$, which we assume is a free particle with momentum operator $\hat p_D$ \cite{solinas2015fulldistribution}. The coupling between the detector and the system is described by the interaction Hamiltonian $\hat H_{SD}(t) =  \alpha (t) ~ \hat p_D ~ \hat H_S(t)$. Here the time-dependent $\alpha(t)$ is chosen so that the detector interacts only twice with the system - once at the start of the process at time $t=0$ and once at the end of the process at time $\tau$, i.e. $\alpha(t)= \lambda [\delta(t) - \delta (t-\tau)]$ with $\lambda$ the coupling strength.
Assuming that the particle mass is so large that its kinetic energy can be neglected and setting $\hbar=1$, the total evolution operator for $S$ and $D$ is 
\begin{equation} \label{eq:V}
	\hat V_{SD,\lambda}(\tau)= e^{i \,  \lambda \, \hat p_D \, \hat H_S(\tau) }  \, \hat V_S(\tau) \,  e^{- i \,  \lambda \, \hat p_D \, \hat H_S(0)},
\end{equation}
where $\hat V_S(\tau) = \Forder \exp \left(-i \int_0^{\tau} \d t \, \hat H_S(t) \right)$ is the system evolution operator for the time-interval $[0, \tau]$, showing that the evolution of the system remains unperturbed \cite{solinas2015fulldistribution, solinas2016probing}.  We will assume that the system and the detector are initially in a product state, $\rho_{SD}(0) = \rho_S (0) \otimes \ket{\phi}_D\bra{\phi}$, where $\rho_S (0)$ is the initial density matrix of the system and the detector is initially prepared in a pure state, $\ket{\phi}_D$. The system and detector dynamics is shown schematically in Fig.~\ref{fig:S+D}.

One can see that when the unitary $\hat V_{SD,\lambda}(\tau)$ acts on a momentum eigenstate $\ket{p}_D$ of the detector, the interaction Hamiltonian $\hat H_{SD}$ will cause a phase shift that is conditioned on the energetic state of the system at the beginning and end of the process. In contrast, when acting on a position eigenstate $\ket{x}_D$ of the detector,  $\hat H_{SD}$ will cause a displacement of the detector conditioned on the energetic state of the system. This observation motivates two complementary methods to extract the information about the work done.

\subsection{Initial detector state and final readout}

We now consider two measurement protocols that differ by their choice of initial detector state $\rho_D(0) = \ket{\phi}_D\bra{\phi}$ \cite{solinas2016probing} and the measurement of the final detector state, $\rho_D(\tau) = \Trace_S[\rho_{SD} (\tau)]$. 

\medskip

{\bf Protocol 1} follows the Full Counting Statistics (FCS) approach \cite{nazarov2003full, Belzig2001, bednorz2010quasiprobabilistic,bednorz2012nonclassical}. Here the detector is initially prepared in an arbitrary superposition of momentum eigenstates, $\ket{\phi}_D = \int \d p \, \phi(p) \, \ket{p}_D$, described by a wavefunction $\phi(p)$ in momentum space. The final detector readout measures the relative phase between different momentum states $\ket{p}_D$ in the final detector state $\rho_D( \tau)$. It has been shown that the phase difference accumulated between eigenstates of detector momentum $\ket{p/2}_D$ and $\ket{-p/2}_D$ is related to the characteristic function of the work, $\GFunc_\lambda$, given by \cite{solinas2015fulldistribution}
\begin{eqnarray}
	\GFunc_\lambda 
	&=& {}_D\matrixel{p/2 }{\rho_D(\tau) }{-p/2}_D/{}_D\matrixel{p/2 }{\rho_D (0) }{-p/2}_D. 
\end{eqnarray} 
This function contains information about all the work moments \cite{mazzola2013measuring, dorner2013extracting, roncaglia2014work, solinas2015fulldistribution} and it can be measured by standard tomographic techniques  \cite{Batalho2014}.
Note that measurement of the function $\GFunc_\lambda$ requires repeated experiments with varying coupling strength, or, alternatively letting the coupling act for varying times. 

Indeed the full work probability distribution can be obtained as the Fourier transform of $\GFunc_\lambda$, $\Prob(W)= \int \d \lambda~\exp{\{ - i  \lambda W \}} \, \GFunc_\lambda$, and it reads \cite{solinas2016probing}  
\begin{equation} \label{eq:quasi_P}
	\Prob(W) =  \sum_{ijk} \, \rho_{ik}^0 \, V_{kj}^\dag \, V_{ji} \, \, \delta 
	\Big[W - \left(\epsilon_j^\tau- \frac{\epsilon_i^0+\epsilon_k^0}{2}\right) \Big].
\end{equation}
Here $\rho_{ik}^0 = \matrixel{\epsilon_i^0}{\rho_{S} (0)}{\epsilon_k^0}$, $V_{ji}=\matrixel{\epsilon_j^\tau}{\hat V_S(\tau)}{\epsilon_i^0}$, and $V_{ji}^\dag=\matrixel{\epsilon_j^0}{\hat V_S^\dag(\tau)}{\epsilon_i^\tau}$ are matrix elements of system operators with respect to the system's  time-dependent energy basis where $\ket{\epsilon_i^t}$ denotes the $i$-th eigenstate of $\hat H_S(t)$ and $\epsilon_i^t$ its corresponding eigenvalue at time $t$.

We refer to the diagonal contributions of the system's initial state $\rho_{S}(0)$ in the energy basis, i.e., $i=k$, as the \emph{classical contributions} since they can be directly attributed to classical transitions between well-defined energetic values, $\epsilon_j^\tau - \epsilon_i^0$. The contributions to the work probability distribution arising from off-diagonal matrix elements,  i.e. $i\neq k$, are associated with classically forbidden energy exchanges of half of the energy gaps, see Eq.~\eqref{eq:quasi_P}. When these off-diagonals are non-zero the work distribution $\Prob(W)$ is not positive-definite and represents a quasi-probability distribution \cite{solinas2015fulldistribution, allahverdyan2014, hofer2016, Perarnau-Llobet2017No-Go,miller2016time, lostaglio2017quantum}. This follows from the fact that the diagonal $i=k$ terms in Eq.~\eqref{eq:quasi_P} sum to unity whilst $\Prob(W)$ always remains normalised. Despite their counterintuitive interpretation quasi-probabilities, such as e.g. the Wigner function, are not new in quantum mechanics and are usually manifestation of pure quantum features of the system \cite{mitchison2007sequential}. In the present case, the negativity of $\Prob(W)$ can be related to the violation of the Leggett-Garg inequality and it is considered a signature of the quantumness of the closed quantum process \cite{bednorz2010quasiprobabilistic,bednorz2012nonclassical, clerk2011full, solinas2015fulldistribution}.

\medskip

In {\bf Protocol 2} the detector is initially prepared in a Gaussian state that approximates the position eigenstate $\ket{x_0}_D$
\begin{equation} \label{eq:Gaussian}
	\tilde{\phi}_{\sigma}(x) = {1 \over (2 \pi \sigma^2)^{1/4}} \, e^{- {(x- x_0)^2 \over 4 \sigma^2}}.
\end{equation}
The variance $\sigma^2$ gives the uncertainty of the initial state and limits the precision of the final measurement. It is also related to the perturbation induced in the system because of the interaction with the detector allowing us to pass from strong (precise) measurement for $\sigma \rightarrow 0$, to weak (imprecise) measurement for $\sigma \rightarrow \infty$ \cite{talkner2016}. The measurement after the evolution reads out the detector's position, i.e. $\rho_D (\tau)$ is measured in the position basis $\{\ket{x}_D\}$. 

The observed position shift can then be related to the work done on the system as $x-x_0 = - \lambda \, W$ \cite{roncaglia2014work,talkner2016, solinas2016probing}.
Assuming the detector is initially centered at $x_0=0$ the probability for the detector to be shifted to position $x$ reads
\begin{equation} \label{eq:P_W}
  	\Prob_{\sigma}(x) = \sum_{ijk}  \, \rho_{ik}^0 \, V_{kj}^\dag \, V_{ji} ~ 
	\tilde{\phi}_{\sigma} \left(x + \lambda \epsilon_{ji}   \right) 
	\tilde{\phi}^*_{\sigma} \left(x + \lambda  \epsilon_{jk} \right). 
\end{equation}
where $\epsilon_{jk} = \epsilon^\tau_{j} - \epsilon^0_{k}$. We note that in Protocol 2 the work distribution $\Prob_{\sigma}(x)$ is always positive since it is defined as ${}_D\matrixel{x}{\rho_D (\tau)}{x}_D$ \cite{solinas2016probing}. Quantum features, such as signatures of initial coherences and interferences, arise when one is unable to distinguish evolutions related to different position shifts (i.e. work values)  \cite{talkner2016, solinas2016probing}. As will be illustrated for an example in Section \ref{sec:example} this occurs when there is substantial overlap between $\tilde{\phi}_{\sigma} \left(x + \lambda \epsilon_{ji}   \right)$ and $\tilde{\phi}_{\sigma} \left(x' + \lambda  \epsilon_{ji} \right)$ for final positions $x$ and $x'$.

\subsection{Initial system state}

We are interested in the effect of the system's initial quantum coherences on the work averages for the two protocols.  For simplicity we will here assume an initial system state that differs from the standard thermal equilibrium state $\rho_S^{eq} = {e^{-\beta \hat H_S(0)} / Z}$, with partition function $Z= \Trace [e^{-\beta \hat H_S(0)}]$, 
by having non-zero initial coherences between eigenstates of $\hat H_S(0)$ \cite{Elouard2015stochastic, Korzekwa2016}. I.e. here we choose the initial system states to be pure, $\rho_S (0) = \ket{\Psi}\bra{\Psi}$, with
\begin{equation} \label{eq:initial}
	\ket{\Psi} = \frac{1}{\sqrt{Z}} \, \sum_k \, e^{i \varphi_k} \, 
	e^{-\beta \epsilon^0_k/2} \, \ket{\epsilon^0_k},
\end{equation}
where $\varphi_k$ are arbitrary phases. The probability of the $k$-th energy state to occur is Boltzmann distributed, i.e. $p_k = |\braket{\epsilon^0_k}{\Psi}|^2 = e^{-\beta \epsilon^0_k}/Z$, and all moments of the energy are the same as the equilibrium state, i.e. $\Trace[ (\hat H_{S}(0))^k \, \rho_S (0)] = \Trace[ (\hat H_{S}(0))^k \rho_S^{eq}]$. Therefore, the pure initial state $\ket{\Psi}$ is energetically indistinguishable from the initial equilibrium state, but it has additional resources related to the presence of coherences \cite{Baumgratz2014QuantCoher, Winter2016TheoCoherences, liu2016theory, Napoli2016, Deffner2016, Chitambar2016Critical, Korzekwa2016, Perarnau-Llobet2017No-Go, Elouard2017, miller2016time, lostaglio2017quantum}. To identify the effect of the initial coherences on the work averages we will also evaluate these averages in the standard thermal equilibrium state $\rho_S^{eq}$.

\section{Jarzynski equality and average work}

The two work distributions (\ref{eq:quasi_P}) and (\ref{eq:P_W}) obtained above, by coupling a detector  to the system with different initial states and performing different measurements, now allow the calculation of any expectation value of the work done on the system. For simplicity we will assume a periodic drive, i.e., $ \hat H_{S}(\tau)= \hat H_{S}(0)$, and denote the energy eigenvalues and eigenstates without the time-superscript as $\epsilon_i$ and $|\epsilon_i\rangle$, respectively. Extensions to non-periodic Hamiltonian dynamics follow straightforwardly. We will here focus on evaluating the exponentiated work averages, $\average{e^{-\beta W}} =  \int \d W  \, e^{-\beta W} \, \Prob(W)$, that will enable us to compare with extensive literature on the quantum Jarzynski equality \cite{campisi2011colloquium,campisi2011erratum,esposito2009Erratum,esposito2009nonequilibrium}, as well as the average work, $\average{W} =  \int \d W \, W  \, \Prob(W)$, which is bounded by the second law in equilibrium thermodynamics.

In the TPM scheme the average exponentiated work for a system that starts initially in a thermalised state, $\rho_S^{eq}$, gives the Jarzynski equality $\average{e^{-\beta W}} = 1$, which is satisfied for any initial inverse temperature $\beta$ and any drive \cite{tasaki2000jarzynski, kurchan2000quantum, talkner2007fluctuation, esposito2009nonequilibrium, campisi2011colloquium}. We note that in the TPM scheme the coherences in $\ket{\Psi}$ would be removed in the first measurement and thus all work averages for processes starting in $\ket{\Psi}$ are identical to those obtained when starting with the thermal state $\rho_S^{eq}$. 

\subsection{Protocol 1: Measuring energy change by phase difference. }

For {\bf Protocol 1} the average exponentiated work for a periodic drive on a system initially in state $\ket{\Psi}$ is obtained with $\Prob(W)$ from Eq.~(\ref{eq:quasi_P}), and can be expressed as
\begin{eqnarray} \label{eq:Generalized_JE_2}
 	\average{e^{-\beta W}} 
	&=&  \sum_{ijk}  \, e^{i (\varphi_i-\varphi_k)}  
	\, \bra{\epsilon_k} \hat V_S^\dag(\tau) \ket{\epsilon_j} \, \frac{e^{-\beta \epsilon_j}}{Z} 
	\, \bra{\epsilon_j}  \hat V_S(\tau) \ket{\epsilon_i}. \quad
\end{eqnarray}
If the initial state was the thermal state $\rho_S^{eq}$ instead, then the absence of off-diagonal terms in  $\rho_S^{eq}$ corresponds to summing only over $k=i$ and $j$ in the sum above, which recovers the JE, $\average{e^{-\beta W}} =  1$. For the pure initial state $\ket{\Psi}$, defining  $\ket{\bar{\epsilon}_i} = e^{i \varphi_i} \, \hat V_S(\tau) \, \ket{\epsilon_i}$ and separating in Eq.~\eqref{eq:Generalized_JE_2} the diagonal terms, $ \sum_{k}  \, \bra{\bar{\epsilon}_k} \, \rho_S^{eq} \, \ket{\bar{\epsilon}_k} = \Trace[ \rho_S^{eq}] =1$, from the off-diagonal terms one obtains
\begin{equation} \label{eq:Generalized_JE_3}
 	\average{e^{-\beta W}} 
	=  1+ \sum_{i \neq k}    \, \bra{\bar{\epsilon}_k} \, \rho_S^{eq} \, \ket{\bar{\epsilon}_i}.
\end{equation}
One can see immediately that in addition to the unit term for an initial thermal state, a term appears that is linked to the off-diagonals of the pure state $\ket{\Psi}$ with respect to the initial Hamiltonian $\hat H_S (0)$. This term is expressed as the sum of the off-diagonals of the canonical state $\rho_S^{eq}$ but with respect to the basis $\{\ket{\bar{\epsilon}_i}\}$, and while the term is real it can  be positive or negative. 

\medskip

Various coherence measures have been proposed by Baumgratz \textit{et al.} \cite{Baumgratz2014QuantCoher}. These measures quantify the amount of coherence in a state $\rho$ that is available as a resource with respect to some fixed basis. For example, one such measure is the $l_{1}-\text{norm}$ of coherence, defined as
\begin{equation}
	C_{l_{1}}(\rho)=\sum_{k\neq i}|\rho_{k, i}|
\end{equation}
where $\rho_{k, i}$ denote the off-diagonal terms of state $\rho$ with respect to the coherence basis $\lbrace \ket{k} \rbrace$. 
Turning to Eq.~\eqref{eq:Generalized_JE_3} and denoting the coherence basis as the evolved energy eigenstates $\lbrace \ket{k} \rbrace \equiv \lbrace  \ket{\bar{\epsilon}_k} \rbrace$, we see that the deviation from the standard Jarzynski equality is bounded by the $l_{1}-\text{norm}$
\begin{eqnarray}
	\left|\average{e^{-\beta W}}  - 1 \right| 
	&=&  \left |\sum_{k \neq i}  \bra{\bar{\epsilon}_k} \, \rho_S^{eq} \, \ket{\bar{\epsilon}_i}  \right| 
	\leq \sum_{k \neq i}   \left |\bra{\bar{\epsilon}_k} \, \rho_S^{eq} \, \ket{\bar{\epsilon}_i}  \right| \nonumber \\
	&=& C_{l_{1}}(\rho_S^{eq}),
\end{eqnarray}
where we have used the triangle inequality.

\medskip

An important consequence of the Jarzynski equality is that it gives a lower bound on the average work done on the system. By using Jensen's inequality, the standard JE for an initial thermal state $\rho^{eq}_S$ implies that the average work done on the system is bounded by the free energy difference, $\average{W} \geq \Delta F$  with $\Delta F =0$ for a periodic drive. Unfortunately, the Jensen inequality cannot be used for Eq.~(\ref{eq:Generalized_JE_2}) because $\Prob(W)$ in Eq. (\ref{eq:quasi_P}) is a quasi-probability and this breaks the convexity property necessary for the Jensen's inequality to be applicable.

Nevertheless, direct calculation of the average work using $\Prob(W)$ from Eq.~(\ref{eq:quasi_P}), gives
\begin{eqnarray}	\label{eq:W_protocol1}
 	\average{W} &=&  \frac{1}{Z} \, 	\sum_{i j}  \, e^{-\beta \epsilon_i}\, \left(\epsilon_j- \epsilon_i\right) \, V_{ij}^\dag \, V_{ji}   \\
	&+&  \frac{1}{Z} \, \sum_{i \neq k,j} \, e^{i (\varphi_k-\varphi_i)} \, e^{-\beta (\epsilon_i + \epsilon_k)/2} \,\left(\epsilon_j - \frac{\epsilon_i +\epsilon_k}{2}\right) \,  V_{kj}^\dag \, V_{ji} . \nonumber
\end{eqnarray} 
The first line represents the classical contribution obtained in absence of initial coherences and it can be shown that it is always positive. This is what one would expect if no initial coherences are present. The terms on the second line are quantum contributions that arise from initial coherences. Interestingly, in the limit $\beta \rightarrow \infty$, i.e. $T \to 0$, these off-diagonal contributions vanish while they are important at all non-zero temperatures. 

The effect of the initial quantum coherence on the  average work can be seen from Eq. (\ref{eq:W_protocol1}).
Once one fixes the dynamics $\hat V_S (\tau)$ and the inverse temperature $\beta$ it still depends on the phase differences $\varphi_k-\varphi_i$. This means that one can modify the average value of the work just by changing the initial state coherences while not affecting the state populations. The example in Section \ref{sec:example} illustrates that $\average{W}$ can also change sign implying  that by controlling the quantum coherences one can extract work from the system instead of exciting it at a work cost.

\subsection{Protocol 2: Measuring energy change by position difference.}

For {\bf Protocol 2} the average exponentiated work for a periodic drive  is obtained with $\Prob_{\sigma}(x)$  from Eq.~(\ref{eq:P_W}), using $x= -  \lambda W$. For the initial system state $\ket{\Psi}$ it can be expressed as
\begin{eqnarray}   \label{eq:JE_protocol_2}
  	 \average{e^{-\beta W}} &&
	=  \sum_{ijk} e^{i \varphi_{ik}}  \, V_{ji} \, V_{kj}^\dagger   \\
 	 \times&&  \int \d W \, \lambda \, \frac{e^{-\beta \left(W +\frac{\epsilon_i+\epsilon_k}{2}\right)}}{Z} \, \tilde{\phi}_{\sigma} \left(\lambda (\epsilon_{ji}-W)   \right) 
	\tilde{\phi}^*_{\sigma} \left(\lambda (\epsilon_{jk}-W) \right).  \nonumber
\end{eqnarray}
One can see that for $i \not = k$ the two Gaussian packets $\tilde{\phi}_{\sigma}$ overlap as long as $|\lambda \epsilon_{jk} - \lambda \epsilon_{ji}| = |\lambda \epsilon_{ik}| \le \sigma$. Phase dependent terms then contribute to the average exponentiated work.
Separating again the diagonal and the off-diagonal terms we obtain  (see Appendix \ref{app:protocol2-expW})
\begin{equation}  \label{eq:avgexpW_prot2}
  	\average{e^{-\beta W}} =   e^{\beta^2 \sigma^2 \over 2 \lambda^2} 
	\left( 1   + \sum_{k\not = i}   \, \bra{\bar \epsilon_k} \, \rho_S^{eq} \,  \ket{\bar \epsilon_i} \, e^{- {\lambda^2 \epsilon_{ik} \over 8 \sigma^2}} \right),
\end{equation}
where $\epsilon_{ik}= \epsilon_i - \epsilon_k$ and the term $e^{- {\lambda^2 \epsilon_{ik} \over 8 \sigma^2}}$ arises through the weighting with the detector's initial state wavefunction, 
\begin{eqnarray} 
  	\int \d W \, \lambda \, e^{-\beta W} \, \tilde{\phi}_{\sigma} \left(\lambda (\epsilon_{ji}-W)   \right) \tilde{\phi}^*_{\sigma} \left(\lambda (\epsilon_{jk}-W) \right) \nonumber \\
	= 	e^{- {\lambda^2 \epsilon_{ik} \over 8 \sigma^2}} \, e^{- \beta \left(\epsilon_j - \frac{\epsilon_i+\epsilon_k}{2}\right)} \, e^{\beta^2 \sigma^2 \over 2 \lambda^2}
\end{eqnarray}  
If the system instead started in the thermal state $\rho_S^{eq}$ the second term in Eq.~\eqref{eq:avgexpW_prot2} vanishes and the average exponentiated work is just given by $e^{\beta^2 \sigma^2 \over 2 \lambda^2}$. 

For either pure or thermal initial system state the dependence of the exponentiated work on the uncertainty in the initial detector position, quantified by $\sigma$, is contained in the pre-factor which diverges exponentially with $\sigma$, as was observed in Ref.~\cite{talkner2016}. Physically this is related to the weakness of the measurement performed, which perturbs the system less than the projective measurement, but at the same time increases the uncertainty of the measurement outcome. Thus the weaker the measurement, i.e. the larger $\sigma$, the larger the second and higher moments of work. Since the average exponentiated work includes information about all moments it hence diverges for infinitesimally weak measurement ($\sigma \rightarrow \infty$) where the detector can be in any position \cite{talkner2016}. 

In the limit that the detector's initial state was a perfect position eigenstate, i.e. $\sigma \rightarrow 0$, the first term in \eqref{eq:avgexpW_prot2} approaches 1 exponentially, as expected. The second term  shows dependence on the system's initial state coherences, given by non-zero $\varphi_{ik}$ and $\epsilon_{ik}$, as well as dependence on coherences that the unitary evolution $\hat V_S$ produces when acting on the diagonal of $\ket{\Psi}\bra{\Psi}$, which is given by $\rho_S^{eq}$. 
The second term in \eqref{eq:avgexpW_prot2} makes the most interesting contribution when the sharpness of the initial detector state is of the same order as the energy difference between two energy eigenvalues of the system, i.e. when $\sigma \approx |\lambda \epsilon_{jk}|$. From a physical point of view this corresponds to the situation in which one cannot distinguish between the position measurement outcomes because the detector uncertainty is so large and interference effects appear \cite{feynman1965quantum, solinas2016probing}. In this situation the initial system coherences will, for moderately small temperatures, make the average $\average{e^{-\beta W}}$ deviate significantly from the thermal state value of $e^{\beta^2 \sigma^2 \over 2 \lambda^2}$ as a function of initial state phases, see Fig.~\ref{fig:protocol2}b) for an example. As soon as $\sigma \ll \epsilon_{ik}$ the $\tilde{\phi}_{\sigma}$ functions in the integral have no overlap (since the system has no degenerate states), the initial coherences have no effect, the second term vanishes and one recovers the thermal state result.

The average work done on the system as measured with {\bf Protocol 2} can be expressed as 
\begin{equation}  
	\average{W} =  \Trace_{S} [(\sigma_S - \rho_S (0)) \,  \hat H_{S}],
\end{equation}
where $\sigma_S = \int \d p \, |\phi_{\sigma} (p)|^2 \, \hat V_{S} (\tau)  \, e^{-i \lambda p \hat H_S} \, \rho_S (0) \, e^{i \lambda p \hat H_S} \, \hat V_{S}^{\dag}(\tau)$ is the time-evolved state of the system, weighted by the initial momentum distribution of the detector. The weighting involves $\phi_{\sigma} (p)$, which is the Fourier transform of $\tilde \phi_{\sigma} (x)$, and depends on the precision of the initial detector state $\sigma$.
When the initial system state is $\rho_S^{eq}$ the coupling to the detector commutes with the state and the average work simplifies to  $	\average{W}^{eq} =  \Trace_{S} [(\hat V_{S} (\tau)  \, \rho_S^{eq} \, \hat V_{S}^{\dag} (\tau) - \rho_S^{eq}) \,  \hat H_{S}]$, as expected. In contrast for the initial pure state the average work becomes  (see Appendix \ref{app:protocol2-W})
\begin{eqnarray} \label{eq:AvgW_prot2}
	\average{W} 
	&=& \average{W}^{eq} +  \sum_{k\not = i} \,   \bra{\bar \epsilon_k} \,  \hat H_{S}  \ket{\bar \epsilon_i}  \, {e^{- \beta {\epsilon_{i} + \epsilon_{k} \over 2}} \over Z} \,   e^{- {\lambda^2 \epsilon_{ik}^2 \over 8 \sigma^2}} 
\end{eqnarray}
which depends on the initial system phases through $\ket{\bar{\epsilon}_i} = e^{i \varphi_i} \, \hat V_S(\tau) \, \ket{\epsilon_i}$.

\section{Example: Driven qubit} \label{sec:example}

To visualise the work associated with coherences we here consider the example of a periodically driven two level system with initial and final Hamiltonian $\hat H_{S} = - (\Delta/2) ~ \sigma_z$ where $\sigma_i$ ($i=x,y,z$) are the Pauli matrices and $\Delta = \epsilon_1 - \epsilon_0$ is the energy gap. The time-dependence of the Hamiltonian $\hat H_S(t)$ for $t \in [0, \tau]$ will induce a unitary transformation on the system. Any such unitary can  be expressed as 
\begin{equation}  \label{eq:generic_U}
	\hat V_S(\tau)= e^{ - i \, \delta \,  \vec{n} \cdot \vec{\sigma} } = \cos \delta -i~\vec{n} \cdot \vec{\sigma} \sin \delta
\end{equation}
where $\vec{n} = \{n_x, n_y, n_z\}$ is a normalized vector. The details of the time-evolution will determine the specific vector $\vec{n}$ as well as the phase $\delta$. 
The initial system state, Eq.~\eqref{eq:initial}, for a qubit is
\begin{equation}
	\ket{\Psi}
	= \frac{1}{\sqrt{Z}} \, 
	\left(e^{+\beta \Delta/4} \, \ket{0} + e^{i \varphi} \, e^{-\beta \Delta/4} \, \ket{1} \right),
\end{equation}
where $\ket{0}$ and $\ket{1}$ are the eigenstates of $\sigma_z$ with eigenvalues $-1$ and $+1$ respectively. The phases have here been chosen as $\varphi_0=0$ and $\varphi_1=\varphi$, without loss of generality. The partition function is $Z = e^{+\beta \Delta/2} + e^{-\beta \Delta/2}$.

\begin{figure}[t]
\includegraphics[trim=0cm 0cm 0cm 0cm, width=0.48\textwidth]{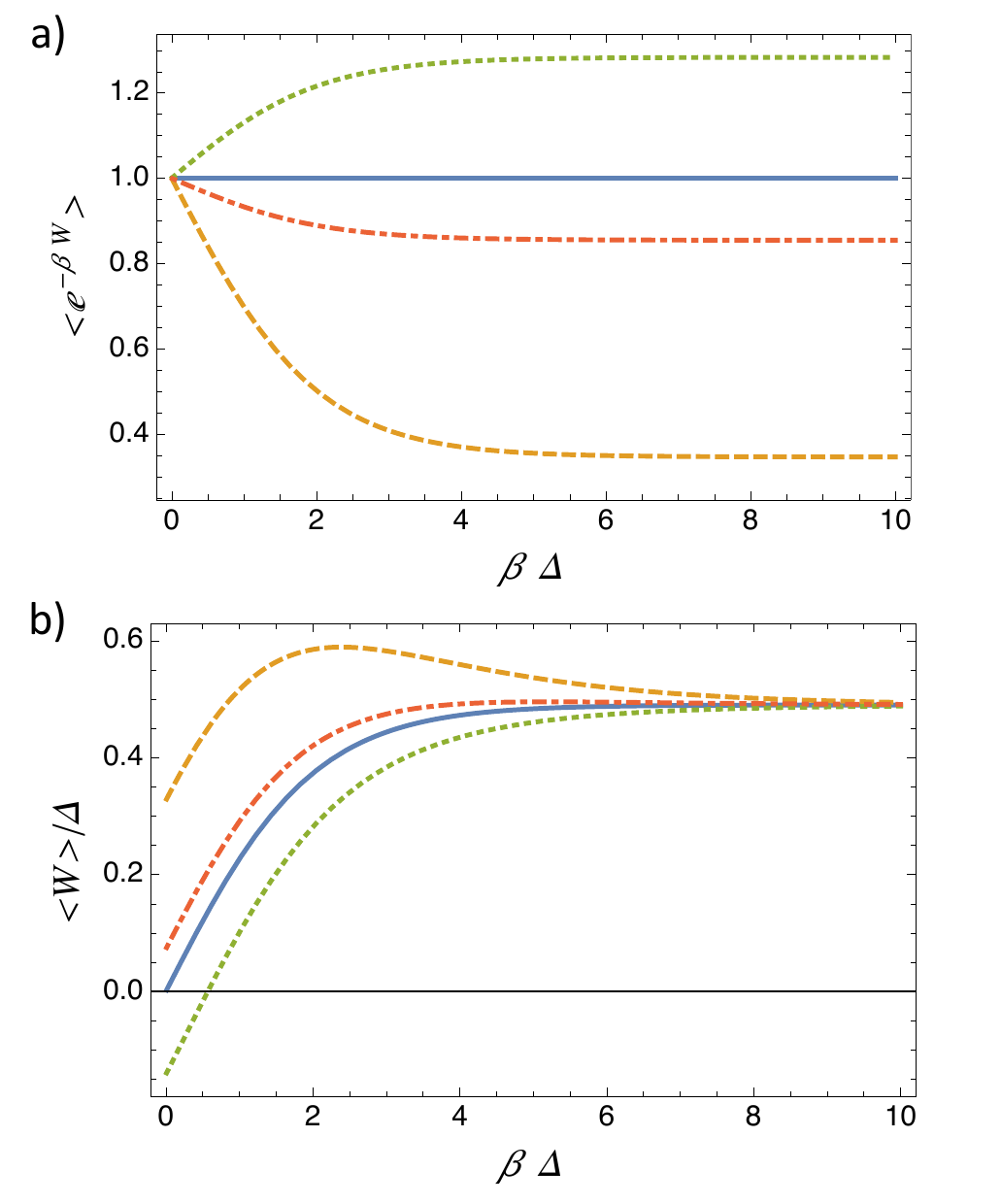}
\caption{
Work measured with the phase measurement ({\bf Protocol 1}) for the qubit example discussed in section \ref{sec:example}.  The unitary transformation $V_S(\tau)$ is characterized by  $\vec{n} =\{ 0.83, 0, 0.55\}$ and $\delta = 1$. The curves are for the initial state $\ket{\Psi}$ with phase $\varphi=4$ (yellow dashed), $\varphi=1$ (red dot-dashed) and $\varphi=0$ (green dotted), respectively.  For comparison, the blue solid curve shows the expectation values when the initial system state was a thermal state $\rho_S^{eq}$. 
{\sf \bfseries a)} Average exponentiated work, Eq.~(\ref{eq:Generalized_JE_3}), over $\beta \Delta$. Deviations from the JE  (blue solid curve) are significant. 
{\sf \bfseries b)} The average work, Eq.~(\ref{eq:W_protocol1}),  for system evolutions starting from initial superposition states $\ket{\Psi}$ deviates from the work for the initial thermal state for small $\beta \Delta$. All curves converge at $\beta \to \infty$, i.e. in the low temperature limit. 
 }
\label{fig:protocol1}
\end{figure}

\subsection{Exponentiated work and average work for Protocol 1.}

The average exponentiated work, Eq.~\eqref{eq:Generalized_JE_3}, and the average work, Eq.~\eqref{eq:W_protocol1}, resulting when the work distribution is measured according to {\bf Protocol 1} are  shown in Fig.~\ref{fig:protocol1} as functions of the inverse temperature $\beta$. The plots show the exponentiated work/work when the system started in the coherent initial states $\ket{\Psi}$ with various phases $\varphi$ and, for comparison, when the initial system state was the thermal state, $\rho_S^{eq}$. In Fig.~\ref{fig:protocol1} {\sf \bfseries a)} deviations of the average exponentiated work from the standard JE value of 1 can be clearly seen for initial states $\ket{\Psi}$ with $\beta > 0$, i.e. for any finite temperature $T \not \to \infty$. 

Fig.~\ref{fig:protocol1} {\sf \bfseries b)} shows the average work done on the qubit for initial states $\ket{\Psi}$ with various phases $\varphi$. For small $\beta$ the initial state coherences do not only ``correct'' the work value but significantly alter it. By controlling the initial phase of $\ket{\Psi}$ one can change $\average{W}$ and even induce an energy emission by the system, i.e. $\average{W}<0$. For example, the yellow-dashed curve for $\ket{\Psi}$ with $\varphi = 4$ has a negative average work for small $\beta$. Thus work is \emph{extracted from} the qubit here in stark contrast to the thermal initial state where work is always \emph{done on} the qubit, indicating that quantum coherences can be used as an energetic resource. 
For large $\beta$, i.e. low temperatures, $\average{W}$ converges to a value independent of the phase $\varphi$, but dependent on the parameters of the unitary Eq.~\eqref{eq:generic_U}. The independence from the phase is clear because at low temperatures the initial superposition state converges to the ground state, $\ket{\Psi} \to \ket{0}\bra{0}$, just like the thermal state $\rho_S^{eq} \to \ket{0}\bra{0}$. In this case, there is only one transition and energy exchange that can occur and the average work becomes this value. 

Maybe surprisingly, the convergence is reversed for $\average{e^{-\beta W}}$. This is because when taking the average of $\average{e^{-\beta W}}$ the exponential suppression is eliminated; indeed that is why $\average{e^{-\beta W}}^{eq}=1$ for any temperature for $\rho_S^{eq}$. However, the coherence effects remain and thus significant corrections to the JE arise from initial state coherences at all finite temperatures. 

\subsection{Exponentiated work and average work for Protocol 2.}

\begin{figure}[t!]
\includegraphics[trim=0cm 0cm 0cm 0cm, width=0.47\textwidth]{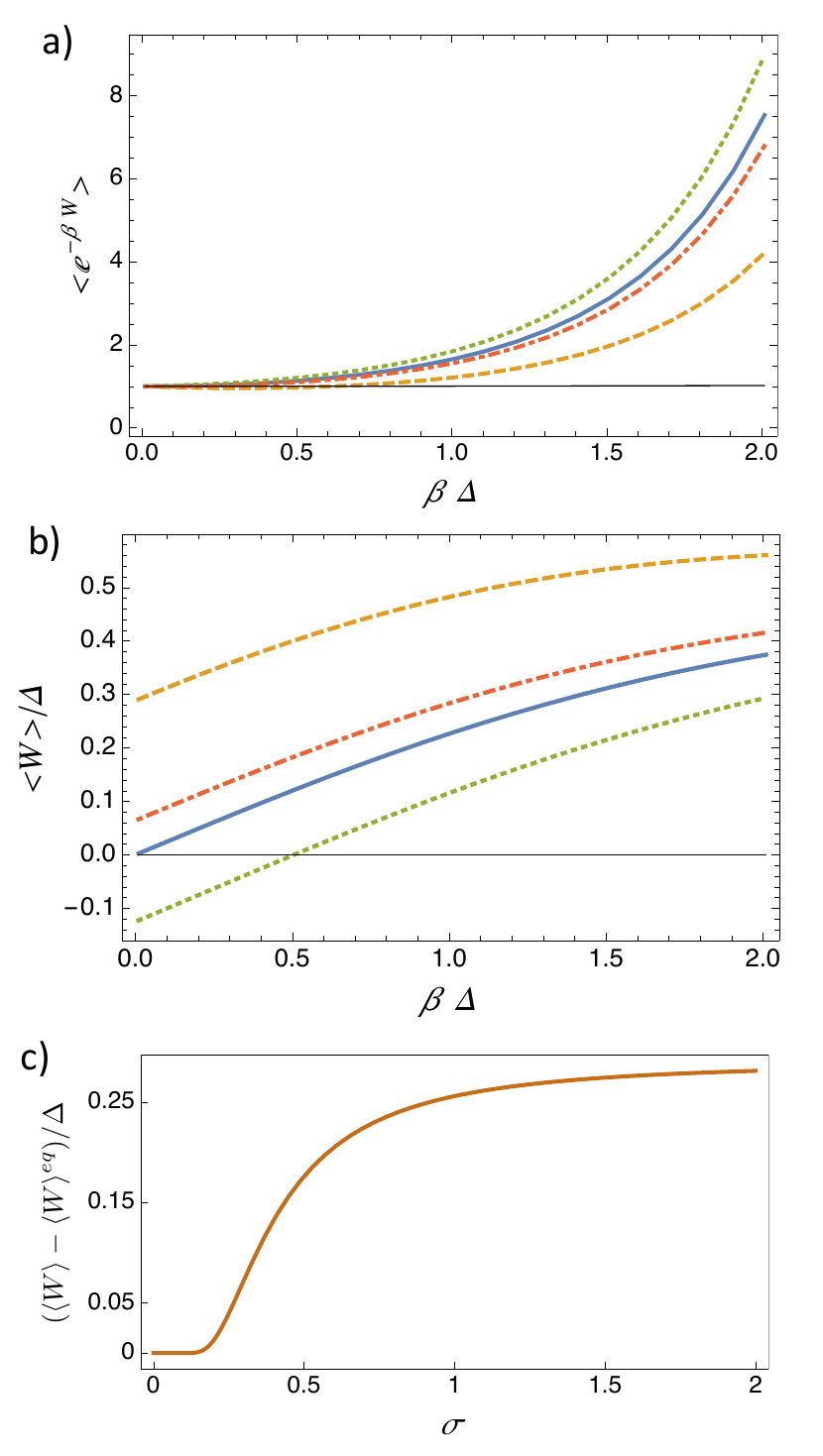}
\caption{\label{fig:protocol2}
Expectation values of work done on the qubit discussed in section \ref{sec:example} when the qubit is weakly measured  using the position of the detector  ({\bf Protocol 2}).  The variance of detector's initial state determined by $\tilde{\phi}_{\sigma}(x)$ is $\sigma = \lambda \Delta $. The unitary transformation $V_S(\tau)$ is characterized by  $\vec{n} =\{ 0.83, 0, 0.55\}$ and $\delta = 1$. Curves shown correspond to the initial state being $\ket{\Psi}$ with phases $\varphi=0$ (yellow dashed), $\varphi=1$ (green dotted) and $\varphi=4$ (red dot-dashed), and for thermal initial state $\rho_S^{eq}$ (solid blue). {\sf \bfseries a)} The average exponentiated work for  initial state  $\ket{\Psi}$ and thermal state $\rho_S^{eq}$ converges to the JE value of 1 at small $\beta \Delta$, indicated by the flat black line. However differences due to initial coherences of the qubit state become pronounced for increasing $\beta \Delta$. 
{\sf \bfseries b)} The average work when the initial state is $\ket{\Psi}$ shows clear dependence on the phase $\varphi$ and differs from the thermal state curve over the whole range of $\beta \Delta$.  
{\sf \bfseries c)} Shown is the difference between the average work done on the system evaluated for initial pure state, $\average{W}$, and initial thermal state, $\average{W}^{eq}$, as a function of $\sigma/\lambda \Delta$. Here $\varphi=0$ is chosen for the pure initial state, and $\beta= {1 / \Delta}$ for both initial state choices.  }
\end{figure}

The average exponentiated work, Eq.~\eqref{eq:avgexpW_prot2}, and the average work, Eq.~\eqref{eq:AvgW_prot2}, resulting when the work distribution is measured according to {\bf Protocol 2} are  shown in Fig.~\ref{fig:protocol2}{\sf \bfseries a)} and {\sf \bfseries b)} as functions of  $\beta \Delta$. The position uncertainty in the initial detector state $\ket{\phi}_D$ is here chosen comparable to the energy gap, $\sigma = \lambda \Delta$. One can see that the average exponentiated work $\average{e^{-\beta W}}$ diverges exponentially with $\beta$ for  $\sigma \neq 0$ as discussed in \cite{talkner2016} even when the system starts initially in the thermal state, $\rho_S^{eq}$. For small $\beta$, i.e. large temperatures, the average exponentiated work for the dynamics starting from an initial superposition state $\ket{\Psi}$ converges to the thermal state value for any initial phase value $\varphi$. However, when $\beta$ is increased, the impact of the coherences are clear: by changing the phase $\varphi$ of the pure state $\ket{\Psi}$ the expectation value $\average{e^{-\beta W}}$ can be tuned to be larger or smaller than the corresponding expectation value calculated with an initial thermal state, $\rho_S^{eq}$. 

Fig.~\ref{fig:protocol2} {\sf \bfseries b)} shows that when the initial detector position uncertainty is $\sigma = \lambda \Delta$ the average work $\average{W}$ done on the system in the unitary protocol increases with $\beta \Delta$ for thermal initial state as well as for superposition states with any phase $\varphi$. When $\beta$ is small,  the initial state coherences have an important effect: tuning the initial phase $\varphi$ drastically changes the average work done  on the system, making it either positive or negative. Thus for some the initial states $\ket{\Psi}$ work can be extracted from the qubit into the drive while for the thermal state the average work is positive for all $\beta \Delta$.

In contrast, if the position uncertainty in the detector's initial state is small, i.e.  $\sigma = 0.1~ \lambda \Delta$, then the initial coherences have no effect on either  $\average{e^{-\beta W}}$  or  $\average{W}$  and the curves in Fig.~\ref{fig:protocol2} {\sf \bfseries a)} and {\sf \bfseries b)} for various $\varphi$ collapse towards the curves for the thermal state $\rho_S^{eq}$. 
Fig.~\ref{fig:protocol2} {\sf \bfseries c)} shows the difference between the average work evaluated for initial pure state, $\average{W}$, and initial thermal state, $\average{W}^{eq}$, as a function of $\sigma/\lambda \Delta$. As expected, for small $\sigma$ the difference of the average works vanishes around $\sigma=0.2 \lambda \Delta$. This is the limit in which the Gaussian functions $\tilde{\phi}_{\sigma}(x+\lambda \epsilon_{ji})$ and $\tilde{\phi}_{\sigma}^*(x+\lambda \epsilon_{jk})$ in \eqref{eq:P_W} are separated and one can distinguish the ``work events'', i.e. excitation or relaxation. As a result the initial system coherences have no effect.  When increasing $\sigma$ the correction arising due to the initial coherence is clearly visible (Fig.~\ref{fig:protocol2} {\sf \bfseries c)}). For sigma much larger than $\lambda \Delta$ the work difference converges quickly to a constant value. This is because for large $\sigma$, energetic transitions cannot be distinguished and the ``work events'' contributing to $\average{W}$ are completely mixed. Increasing the uncertainty $\sigma$ further does not change this and thus the difference between $\average{W}$ and  $\average{W}^{eq}$ plateaus. (This would be different for a multi-state/multi-transition system).

\section{Conclusions}

We have discussed the importance of the initial coherences in the work performed on a quantum system driven by an external classical field. In the common TPM protocol \cite{campisi2011colloquium,campisi2011erratum,esposito2009Erratum,esposito2009nonequilibrium,kurchan2000quantum,tasaki2000jarzynski}, these coherences are destroyed  due to the initial measurement. As a consequence, the system loses much of its quantum features and the work distribution is essentially a classical stochastic one. To preserve the quantum effects, we exploit two alternative measurement protocols that keep track of the quantum features of the system and the dynamics. These protocols \cite{solinas2016probing} are based on weak measurements so that, by changing the implementation procedure, we can control the perturbation induced in the quantum system by the interaction with the detector. 

To quantify the effects of the coherences on the work distribution we calculated $\average{e^{-\beta W}}$ for an initially thermalized system and a system with the same average state populations but with coherences between the eigenstates of the initial Hamiltonian. These two initial states are energetically indistinguishable, i.e., they have the same diagonal distribution in the Hamiltonian basis, so they allow for an immediate identification of the effect of the coherences. The comparison is done taking into account different protocols that can be used to measure the work distributions \cite{solinas2015fulldistribution,solinas2016probing}. While for an initial thermal state we obtain $\average{e^{-\beta W}}=1$, the presence of initial coherences leads to a violation of JE. The degree of violation can be controlled by changing the phase difference between the eigenstates of the initial Hamiltonian that now becomes an experimentally controllable parameter. 

In a similar way, the average work performed on the quantum system depends on the initial coherences. Again, we find that, by changing the initial phase difference between energy eigenstates, we can change the work done on the system. Interestingly, while for periodic drive, the standard initial thermal state will always result in positive work, i.e. the system absorb energy, the presence of initial coherences allows for negative minimal work, i.e. the system emits energy. We note that this does not conflict with the standard second law of thermodynamics, as the initial state \eqref{eq:initial} is not a true thermal equilibrium state. 

The two protocols discussed allow us to extract information about the work statistics of the system; but they are conceptually different. In {\bf Protocol 1}, the physical observable is the accumulated phase in the detector while in {\bf Protocol 2} we measure the detector shift in position. These methods resemble the Full Counting statistics approach \cite{nazarov2003full} and the more traditional weak measurement approach \cite{Aharonov1988weak}, respectively.
The differences between them are exemplifies by the fact that we obtain a quasi-probability distribution for the former and a probability distribution for the work for the latter measurement protocol.  In hind view this should not be so surprising. Since the quantum work is not a standard quantum observable \cite{talkner2007fluctuation,solinas2015fulldistribution,solinas2016probing, talkner2016}, i.e. it is not associable to a hermitian operator, we must specify the details of how we perform its measurement. Different measurement procedures perturb the system in different ways and lead to different measurement outcomes. This implies that one must choose the work measurement and distribution associated with it that suits the task one wants to characterise. 

It is interesting to contrast the work statistics of the two weak measurement protocols with the work statistics arising in the strong TPM protocol. In {\bf Protocol 1},  independently of the strength of the system-detector coupling, one always obtains a quasi-probability distribution; thus the quantum effects are always present. However, in {\bf Protocol 2} one observes convergence to the TPM averages when the measurement uncertainty $\sigma$ is small thus forcing the quantum system to behave classically. Only when the measurement uncertainty $\sigma$ is large, and one cannot distinguish different energetic transitions, the underlying quantum features of the system are exposed. The work distribution and work averages then deviate from the classical TPM results (as shown in Fig. \ref{fig:protocol2}c). The two protocols discussed can be implemented experimentally, for example, using Nuclear Magnetic Resonance (NMR) where the system is a spin within a molecule and an auxiliary spin in the same molecule can play the role of the detector  \cite{Batalho2014}, or using circuit quantum electrodynamics as discussed in \cite{solinas2016probing} where the detector is a bosonic mode.

Coherence is a central feature of quantum mechanics and a key ingredient for quantum technologies, such as quantum cryptography. Our results lead to the conclusion that coherences can be used as an energetic resource: the system's capability to store or emit energy can be modified by simply changing the initial phases of the system. This aspect has been underestimated for a long time. Implications go beyond mere academic interest; but whether the coherence-dependent energetics of the system can be used for quantum technologies remains an open question.

\begin{acknowledgments}
PS has received funding from the European Union FP7/2007-2013 under REA
grant agreement no 630925 -- COHEAT and from MIUR-FIRB2013 -- Project Coca (Grant
No.~RBFR1379UX). HM is supported by EPSRC through a Doctoral Training Grant. JA acknowledges support from EPSRC, grant EP/M009165/1, and the Royal Society. This research was supported by the COST network MP1209 ``Thermodynamics in the quantum regime''.
\end{acknowledgments}

%

\appendix

\section{Derivation of Eq.~\eqref{eq:avgexpW_prot2} for Protocol 2.} \label{app:protocol2-expW}

For the initial system state $\ket{\Psi}$ the average exponentiated work for a periodic drive is 
\begin{eqnarray} 
  	 \average{e^{-\beta W}} &&
	=  \sum_{ijk} e^{i \varphi_{ik}}  \, V_{ji} \, V_{kj}^\dagger   \\
 	 \times&&  \int \d W \, \lambda \, \frac{e^{-\beta \left(W +\frac{\epsilon_i+\epsilon_k}{2}\right)}}{Z} \, \tilde{\phi}_{\sigma} \left(\lambda (\epsilon_{ji}-W)   \right) 
	\tilde{\phi}^*_{\sigma} \left(\lambda (\epsilon_{jk}-W) \right) \nonumber \\ 
	&=&  \sum_{ijk} e^{i \varphi_{ik}}  \, V_{ji} \, V_{kj}^\dagger  \, \frac{e^{-\beta \left(\frac{\epsilon_i+\epsilon_k}{2}\right)}}{Z} \, e^{- {\lambda^2 \epsilon_{ik} \over 8 \sigma^2}} \, e^{- \beta \left(\epsilon_j - \frac{\epsilon_i+\epsilon_k}{2}\right)} \, e^{\beta^2 \sigma^2 \over 2 \lambda^2} \nonumber \\   	
	&=&  \sum_{ik} e^{i \varphi_{ik}}  \, \bra{ \epsilon_k}  V_{S}^\dagger \, \rho_S^{eq}\, V_S \ket{ \epsilon_i} \, e^{- {\lambda^2 \epsilon_{ik} \over 8 \sigma^2}} \,  \, e^{\beta^2 \sigma^2 \over 2 \lambda^2} \nonumber \\   	
	&=&   e^{\beta^2 \sigma^2 \over 2 \lambda^2} 
	\left( 1   + \sum_{k\not = i}   \, \bra{\bar \epsilon_k} \, \rho_S^{eq} \,  \ket{\bar \epsilon_i} \, e^{- {\lambda^2 \epsilon_{ik} \over 8 \sigma^2}} \right),
\end{eqnarray}
where we have integrated over the Gaussians
\begin{eqnarray} 
  	\int \d W \, \lambda \, e^{-\beta W} \, \tilde{\phi}_{\sigma} \left(\lambda (\epsilon_{ji}-W)   \right) \tilde{\phi}^*_{\sigma} \left(\lambda (\epsilon_{jk}-W) \right) \\
	= 	e^{- {\lambda^2 \epsilon_{ik} \over 8 \sigma^2}} \, e^{- \beta \left(\epsilon_j - \frac{\epsilon_i+\epsilon_k}{2}\right)} \, e^{\beta^2 \sigma^2 \over 2 \lambda^2}
\end{eqnarray}  
and used the definitions $V_{ji}=\matrixel{\epsilon_j}{\hat V_S}{\epsilon_i}$, $V_{ji}^\dag=\matrixel{\epsilon_j}{\hat V_S^\dag}{\epsilon_i}$ and $\ket{\bar{\epsilon}_i} = e^{i \varphi_i} \, \hat V_S(\tau) \, \ket{\epsilon_i}$. 

\section{Derivation of Eq.~\eqref{eq:AvgW_prot2} for Protocol 2.} \label{app:protocol2-W}

For the initial system state $\ket{\Psi}$ the average work for a periodic drive is 
\begin{eqnarray} 
  	 \average{W} 
	&=&  \sum_{ijk} e^{i \varphi_{ik}}  \, V_{ji} \, V_{kj}^\dagger  \, \frac{e^{-\beta \left(\frac{\epsilon_i+\epsilon_k}{2}\right)}}{Z} \, \\
 	 \times&&  \int \d W \, \lambda \, W \,  \tilde{\phi}_{\sigma} \left(\lambda (\epsilon_{ji}-W)   \right) 
	\tilde{\phi}^*_{\sigma} \left(\lambda (\epsilon_{jk}-W) \right) \nonumber \\ 
	&=&  \sum_{ijk} e^{i \varphi_{ik}}  \, V_{ji} \, V_{kj}^\dagger  \, \frac{e^{-\beta \left(\frac{\epsilon_i+\epsilon_k}{2}\right)}}{Z} \, \\
 	 \times&&  \int \d W \, {\lambda \over \sqrt{2 \pi \sigma^2}}\, W \, e^{- {\lambda^2 \over 2 \sigma^2} \left( W - \epsilon_j + \frac{\epsilon_i+\epsilon_k}{2}\right)^2} \, e^{- {\lambda^2 \over 8 \sigma^2} \epsilon_{ik}^2}    \nonumber \\ 
	&=&  \sum_{ijk} e^{i \varphi_{ik}}  \, V_{ji} \, V_{kj}^\dagger  \, \frac{e^{-\beta \left(\frac{\epsilon_i+\epsilon_k}{2}\right)}}{Z}  \, e^{- {\lambda^2 \over 8 \sigma^2} \epsilon_{ik}^2}   \, {\lambda \over \sqrt{2 \pi \sigma^2}} \,\\
 	 \times&&  \int \d y \,  (y +  \epsilon_j - \frac{\epsilon_i+\epsilon_k}{2}) \, e^{- {\lambda^2 \over 2 \sigma^2} y^2 }.  \nonumber
\end{eqnarray}
The exponentials are even in $y$ and thus only the linear term in $y$ in the integral and we obtain
\begin{eqnarray}
  	 \average{W} 
	&=&  \sum_{ijk} e^{i \varphi_{ik}}  \, V_{ji} \, V_{kj}^\dagger  \, \frac{e^{-\beta \left(\frac{\epsilon_i+\epsilon_k}{2}\right)}}{Z}  \, e^{- {\lambda^2 \over 8 \sigma^2} \epsilon_{ik}^2}   \, {\lambda \over \sqrt{2 \pi \sigma^2}} \, \\
 	 \times&& (\epsilon_j - \frac{\epsilon_i+\epsilon_k}{2}) \,  \sqrt{2 \pi \sigma^2 \over \lambda^2}  \nonumber 	 \\ 
&=&  \sum_{ik} e^{i \varphi_{ik}}  \, \,  \left(\bra{ \epsilon_k}  V_{S}^\dagger  \hat H_S  V_{S}\ket{ \epsilon_i}  -  \braket{ \epsilon_k}{ \epsilon_i} \frac{\epsilon_i+\epsilon_k}{2} \right) \\
 	 \times&& \, \, \frac{e^{-\beta \left(\frac{\epsilon_i+\epsilon_k}{2}\right)}}{Z}  \, e^{- {\lambda^2 \over 8 \sigma^2} \epsilon_{ik}^2}.   \nonumber 
\end{eqnarray}
Considering only contributions from the initial system's state diagonal, which has thermal weights just like $\rho_S^{eq}$, this expression reduces to
\begin{eqnarray}
	\average{W}^{eq} 
	&=&  \sum_{i}  \,  \left(\bra{ \epsilon_i}  V_{S}^\dagger  \hat H_S  V_{S}\ket{ \epsilon_i}  -  \epsilon_i \right)  \, \frac{e^{-\beta \epsilon_i}}{Z}   \\
	&=& \Trace[ (V_{S}^\dagger  \hat H_S  V_{S} - \hat H_S) \, \rho_S^{eq}].
\end{eqnarray}
Thus for the pure initial state $\ket{\Psi}$ the average work becomes
\begin{eqnarray}
	\average{W}
	&=& \average{W}^{eq} +  \sum_{k\not = i} \,   \bra{\bar \epsilon_k} \,  \hat H_{S}  \ket{\bar \epsilon_i}  \, {e^{- \beta {\epsilon_{i} + \epsilon_{k} \over 2}} \over Z} \,   e^{- {\lambda^2 \epsilon_{ik}^2 \over 8 \sigma^2}}.
\end{eqnarray}

\bigskip

\end{document}